# Performance Estimation of 2*3 MIMO-MC-CDMA using Convolution Code

Mr. Atul Singh Kushwah & Mr. Sachin Manglasheril

*Asst. Professor & Digital Communication & Indore Institute of Science & Technology-II, Indore (M.P), India*

**Abstract** - **In this paper we estimate the performance of 2*3 MIMO-MC-CDMA system using convolution code in MATLAB which highly reduces BER by increasing the efficiency of system. MIMO and MC-CDMA system combination is used to reduce bit error rate and also for forming a new system called MC-CDMA which is multi user and multiple access schemes used to increase the performance of the system. MC-CDMA system is a narrowband flat fading in nature which converts frequency selective to numerous narrowband flat fading multiple parallel sub-carriers to increase the efficiency of the system. Now this MC-CDMA system can also be enhanced by grouping with 2*3 MIMO system which utilizes ZF decoder at the receiver to decrease BER in which ½ rate convolutionally encoded Alamouti STBC block code is used for channel encoding scheme as transmit diversity of MIMO with multiple transmit antenna. And convolution encoder is also used as source encoder or FEC encoder in MIMO-MC—CDMA. Advantage of using MIMO-MC-CDMA using convolution code is due to reduce complexity of system and also for reducing BER and finally to increase gain of system. Now after this we examine system in various modulation techniques like, 8-PSK, 16-QAM, QPSK, 32-QAM, 8-QAM and 64-QAM using MATLAB in Rayleigh fading channel.**

## I. INTRODUCTION

Due to recent requirement of technology for high data rate and reduced probability of error in this paper we combine systems like, CDMA, OFDM and MIMO, which results advanced technique for decreasing bit error rate. MC-CDMA is formed by OFDM and CDMA combination which is multiple access and multi-carrier system [11]. MC-CDMA is narrowband flat fading channel. The MC-CDMA increases the efficiency of wireless communication system by high data rate and low error probability. We also use convolution code as channel encoder or as FEC encoder in this system for decreasing bit error rate.

In this paper we combine MC-CDMA with MIMO to further increase the performance. We use 2*3 MIMO antenna diversity technique which is also called is multiple antenna system in which two transmit antennas and three receive antennas are used and for detection of receive diversity and transmit diversity we utilizes half-rate convolutionally encoded Alamouti STBC code which is also used for the synchronization of system to decrease Inter Symbol Interference. For orthogonality detection of signal Zero-Forcing(ZF) detection scheme is used. Finally MIMO-MC-CDMA [4] is formed by combination of above discussed system and this combination is done by using MATLAB then performance estimation is done by using different modulation techniques like 8-PSK, 16-QAM, QPSK, 32-QAM, 8-QAM and 64-QAM in Rayleigh fading channel.

## II. MULTI-CARRIER CDMA

MC-CDMA [7] is the combination of CDMA and OFDM, results better frequency diversity and high data rates. In MC-CDMA, every symbol is spreaded by code chips and transmitted by several subcarriers. It is not necessary that the number of carriers to be equal to the code length so providing a degree of flexibility in our design. In MC-DS-CDMA data is spreaded in time domain rather than frequency domain. In MC-CDMA single data symbol is transmitted over independent subcarriers. The important advantage of MC-CDMA is enhancing the bandwidth efficiency due to multiple accesses is possible through proper system design by orthogonal codes.

### A. Need of MC-CDMA

MC-CDMA use advantage of both the technique OFDM and CDMA and makes an efficient transmission system by the spreading of input data symbols by spreading codes in frequency domain. The number of narrowband orthogonal subcarriers with symbol period longer than the delay spread and all the subcarriers are affected by same deep fades of the channel at the same time cause increase in performance. Increase in the number of path will increase the performance of system this is increased by mainly two first due to diversity, then, it deteriorate due to the increasing the interference from large number of paths at all users. In general, there are an efficient number of paths that depends upon the system to be used and the number of users. Interference increased as the number of users is increased through all the paths. So, the optimum number of paths decreased.

### B. MCCDMA System Model

MC-CDMA [3,5,6] transmitter and OFDM transmitter have minor difference. In OFDM numerous symbols are transmitted through subcarriers but in MC-CDMA identical symbol is transmitted by unlike subcarriers. The block diagram of MC-CDMA is shown in fig.1. The input information symbols are converted into parallel streams of information. Then each parallel stream is spread using spreading codes in ½ rate convolutional codes.

The OFDM system linked with the CDMA system converts the symbols to time domain by Inverse Fast Fourier Transform and assigns subcarrier used for each symbol. Then the subcarriers are multiplexed to shape as a serial stream of data. Before the transmitting the serial stream that is converted into blocks and every block is divided by a guard frame. The guard frame can be zero symbols or known padding symbols. In OFDM the cyclic prefix are used as guard symbols which have various advantages to remove ISI and the inter-carrier interference (ICI) caused by multipath fading. And hence the cyclic prefix length in such a manner that it is greater than the delay spread of the channel. In MC-CDMA transmission, is frequency non selective fading over each sub carrier. So, if the unique symbol rate is high enough to turn out to be frequency selective fading, the input data comprise to be serial to parallel (S/P) converted into parallel information sequences and each S/P output is multiplied by the spreading code of different length. For improving the performance of the system, a suitable approach for channel estimation is, to utilize dedicated pilot symbols that are inserted periodically in the transmission, also called as block-type pilot channel estimation.

In MCCDMA receiver configuration intended for the jth user is shown in Fig.1. The received signal is primarily down converted. Then, remove the cyclic prefix and the remaining samples are converted serial to parallel to obtain the subcarriers components. The subcarriers are first demodulated by a fast FFT and then multiplied by the gain to merge the received signal energy scattered in the frequency domain.

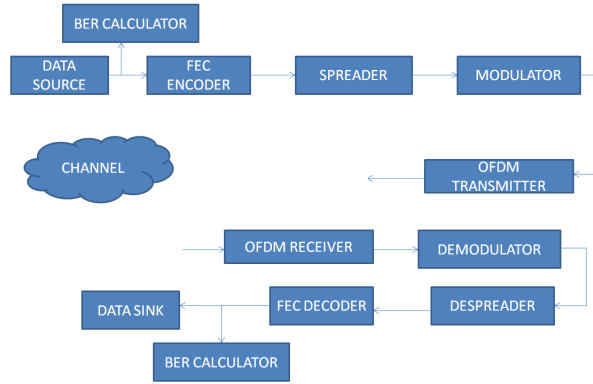

Fig.1 : Block Diagram of MC-CDMA

### III. MIMO OVERVIEW

Multiple antennas are used in MIMO [2] systems at both transmitter and receiver and both transmit and receive diversity are functional to diminish fading ensuing from signal fluctuations all the way through the wireless channel. The system provides diversity gains which is based on the degree at which the multiple data replica be faded independently, represents the difference in SNR at the output of diversity combiner as compared to single branch diversity by the side of certain probability level. MIMO system contains N number of transmit antenna elements equal to two, and of M number of receive antenna elements equal to three was modeled, accordingly diversity order of 6 can be achieved. Combining the several versions of the signals formed by different diversity schemes is desirable for improving the performance. The paper applies zero forcing (ZF) technique decoder to combine M received signals to reverberate on the most probable transmitted signal. The amount of the received SNRs from these M different paths is the efficient received SNR of the system with diversity order M. The receiver requirements is to demodulate all M received signals in case of ZF for a basis with M independent signals in the receiver antennas.

## IV. CONVOLUTION CODE

Non-systematic convolutional encoder [12] is depicted in fig.2. The data were absent at the output of the encoder along with were replaced by modulo 2 sum of the data at that instant is represented by i is $d_i$ and the data at instants $i-2$ is $d_{i-2}$ and so on. The rate of the encoder is ½ i.e. 2 output bits for each input bits.

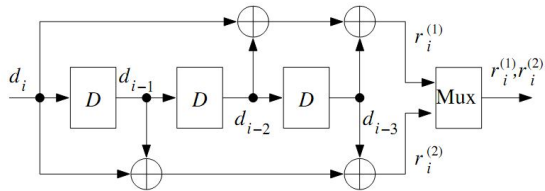

Fig. 2. Non-systematic convolutional encoder [12].

### A.  Generic representation

For the construction of non-systematic convolution encoder shift registers, XOR gates and D flip-flops are used. Now we define the major characteristic of convolution codes that is constraint length, here equivalent to v +1. The register at time i is characterized by the v no. of bits i.e $s_i^{(1)}, s_i^{(2)}, ..., s_i^{(v)}$ then states of v bits are also represented by vector form $s_i = (s_i^{(1)}, s_i^{(2)}, \cdots, s_i^{(v)})$. There are $2^v$ probable state values in convolutional encoder so that we can often denote in binary or binary decimal form. So the state of encoder depends on the no. of flip flops which is determined by $2^3$=8. Let $s_1$ =1, $s_2$ =1 and $s_3$=0, then the state of encoder is 110 in natural binary, i.e. in decimal form is 6.

If we are using m no. of coefficients $a_j^{(1)}$ which is used to form a vector $d_i$ is chosen then addition is done by means of the previous content of flip-flops except not in case of the first flip-flop, which outline the value to be stored in the subsequent flip-flops. New value of a flip-flop depends upon the current input and previous value of flip-flop. If $b_j$ is null coefficients, the consequential input depends upon the sum of the components selected for $d_i$. In case when $b_j$ having non-zero coefficients then these components are sum up with $d_i$ and the repetitive code is generated. So the consecutive states of the registers depends on prior inputs and present input throughout the flip-flop which finally produces many components $r_i$ which is formed by the addition of the content of flip-flops by coefficients g.

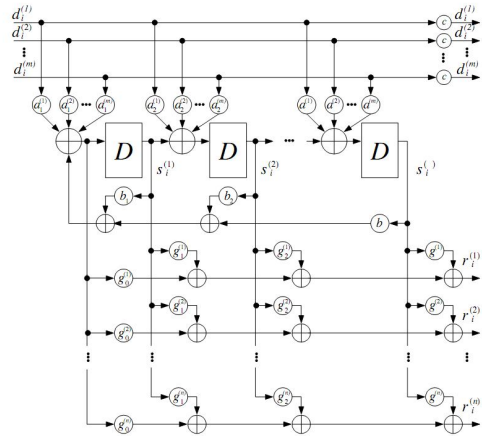

Fig.3. Generic representation of an encoder for convolutional codes [12].

Here we are using non-systematic half-rate convolutional encoder as FEC at the MIMO-MC-CDMA in which forms encoded message bits.

## V.  MIMO-MC-CDMA USING CONVOLUTION CODE SYSTEM MODEL

Communication system model of MIMO-MC-CDMA [1] using convolution code is represented by fig.4.

In this communication system we are assuming that transmitter sends random string to the receiver so we used random PN sequence generator using MATLAB. Now spreading of input sequence is done by PN sequence generator for representing random data input by users. Then in modulator is used for different modulation schemes are used like *8-QAM, 8-PSK, 16-QAM, 32-QAM, QPSK and 64-QAM* this is depicted by modulator block. MC-CDMA system is already explained in section II by Multi-Carrier Code Division Multiple Access. Now MIMO with half rate convolutionally encoded Alamouti's STBC block code is used which is explained in section III as Multiple Input Multiple Output (MIMO). The combination of MC-CDMA and MIMO shape the enhanced system model MIMO-MC-CDMA as shown in fig.4. Now signal is transmitted through Rayleigh Fading Channel [7].

Now receiver obtain the signal in reverse manner with ZF decoder for the revival of transmitted signal at the receiver and BER calculation is done for estimating the system performance. In MIMO two transmit and three receive antennas are used. In this paper we are sending the message bits which are random in nature or data dependent on user then the data is passed through the spreader using PN sequence generator which forms 8 bits for each of the input bits that input bits then resulting bits are formed after the spreading of encoded data sequence. Then these spreading sequences are passed through modulator in which its modulation depends on the different type of modulation to be used. These modulated data is again framed into parallel structure for OFDM then IFFT is done to convert frequency selective carriers into parallel narrowband flat-fading carriers are orthogonal in nature, then this data is converted into parallel to serial then CP cyclic prefix is added to remove ISI which complete the development of OFDM transmission system, then this serial data passed through MIMO with Alamouti STBC code for 2 transmit and 3 receive diversity antenna in which 3*2 channel matrix is formed by using MIMO diversity, also the ZF detection scheme is used at the receiver to detect orthogonality then reverse process is done for receiving the input bits.

## VI. SIMULATION RESULTS AND DISCUSSION

Table 1 represents input model parameters for MIMO-MC-CDMA using convolution code [7,8,9,10] in different modulation technique.

Fig.5 shows the comparative estimation of different modulation schemes in MIMO-MC-CDMA using convolution code.

Table 2 shows the performance analysis of different modulation schemes in terms of gain and BER.

TABLE I

**SIMULATED MODEL PARAMETERS**

| | |
|---|---|
| Channel Encoder | ½ rate convolution encoder Alamouti STBC |
| Signal detection scheme | Zero forcing |
| Signal to Noise Ratio | -10dB to 20 dB |
| CP Length | 1280 |
| OFDM Sub-carriers | 6400 |
| No. of transmitting and receiving antennas | 2*3 |
| Modulation Schemes | QPSK, 8-PSK, 8-QAM, 16-QAM, 32-QAM and 64 QAM |
| Channel | Rayleigh Fading Channel |
| Source Encoder | Half-rate convolution code |

From table.2 and Fig.5 we can see that QPSK shows high gain of 14.04 dB with low BER with respect to different modulation schemes at -4dB SNR. This is done by using MIMO-MC-CDMA using convolution code system by which error probability in QPSK is zero which shows very low probability of error in system.

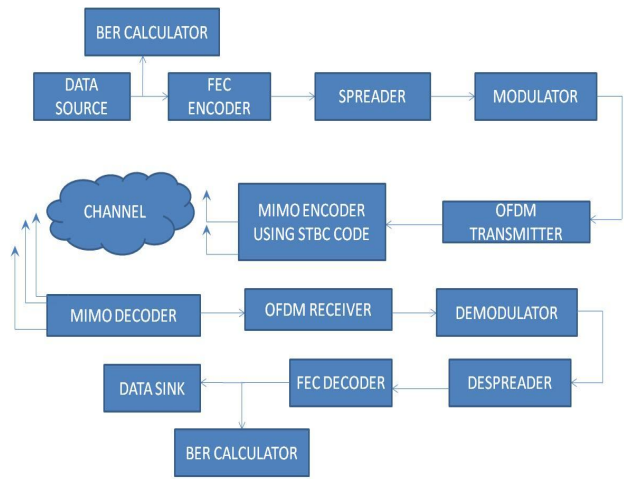

Fig.4. Communication System Model OF 2*3 MIMO-MC-CDMA

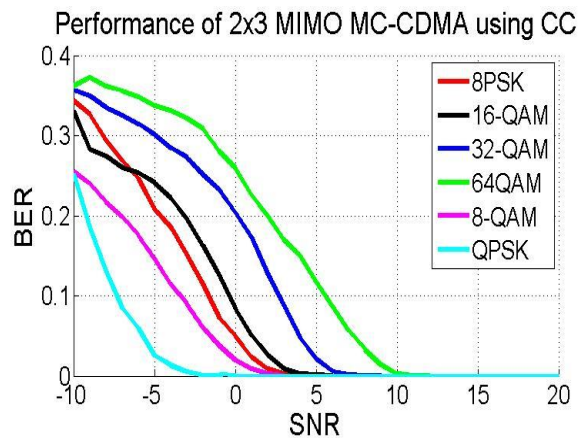

Fig.5. Performance analysis of 2*3 MIMO-MC-CDMA in 8-QAM, 16-QAM, 32-QAM, 64-QAM, 8-PSK and QPSK modulation scheme.

TABLE II

PERFORMANCE ANALYSIS AT -4dB SNR WITH RESPECT TO 64-QAM MODULATION TECHNIQUE AS SHOWN IN FIG.5

| Modulation | BER at -4dB | Gain w.r.t 64-QAM |
|------------|-------------|-------------------|
| QPSK | 0.01308 | 14.04 dB |
| 8-QAM | 0.01144 | 14.62 dB |
| 8-PSK | 0.1859 | 2.51 dB |
| 16-QAM | 0.2221 | 1.736 dB |
| 32-QAM | 0.2843 | 0.6644 dB |
| 64-QAM | 0.3313 | 0dB |

## VII.  CONCLUSION

Fig.5 depicts the comparative estimation of MIMO-MC-CDMA using convolution code in different modulation techniques. Table 2 shows the comparative analysis used for different modulation schemes shows that as modulation order is increased results increase in BER. This manuscript aims to reduced bit error rate which is shown by QPSK modulation scheme as resultant gain of 12.23 dB with respect to 64-QAM modulation technique which shows that the gain of QPSK is highest in comparison to other modulation technique with very low probability of error because errors are finished at 0dB in QPSK modulation. For wireless communication 64-QAM modulation scheme is preferred which contain BER up to 10dB, i.e. errors are remain in 64-QAM upto 10dB SNR which is enhanced by using MIMO-MC-CDMA.